\documentclass[12pt]{article}
\usepackage{amsmath}
\usepackage[dvips]{graphicx}
\usepackage{epsfig}
\usepackage{amsmath}
\usepackage{amssymb}
\usepackage{graphics,epstopdf}
\usepackage{amsfonts}
\usepackage{amsthm}
\usepackage[usenames]{color}

\definecolor{AV}{rgb}{0.65,0.0,0}
\definecolor{GC}{rgb}{0,0.0,0.65}
\definecolor{WS}{rgb}{0,0.65,0}

\usepackage[
      colorlinks=true,
      linkcolor=blue,
      urlcolor=blue,
      filecolor=blue,
      citecolor=red,
      pdfstartview=FitV,
      pdftitle={},
      pdfauthor={},
      pdfsubject={},
      pdfkeywords={},
      pdfpagemode=None,
      bookmarksopen=true
]{hyperref}

\newcommand{\bm}{\begin{multiline}}
\newcommand{\beq}{\begin{equation}}
\newcommand{\eeq}{\end{equation}}
\newcommand{\beqs}{\begin{eqnarray}}
\newcommand{\eeqs}{\end{eqnarray}}

\begin{document}

\title{\textbf{\Large Charged particles in the background of the Kiselev solution in power-Maxwell electrodynamics}}

\author{\textbf{Marina--Aura Dariescu}$^1$\footnote{E-mail: marina@uaic.ro}, \textbf{Vitalie Lungu}$^2$\footnote{E-mail: vitalie.lungu@student.uaic.ro}, \textbf{Ciprian Dariescu}$^3$\footnote{E-mail: ciprian.dariescu@uaic.ro} \\  \textbf{and Cristian Stelea}$^{4}$\footnote{Corresponding author e-mail: \texttt{cristian.stelea@uaic.ro}} \\
$^{1,2,3}$Faculty of Physics, 
{ ``Alexandru Ioan Cuza"} University  of Iasi \\
Bd. Carol I, No. 11, 700506 Iasi, Romania \\
$^{4}$Department of Exact and Natural Sciences, \\
Institute of Interdisciplinary Research,\\
 { ``Alexandru Ioan Cuza"} University of Iasi, \\
 Bd. Carol I, no. 11, 700506 Iasi, Romania 
}

\date{}
\maketitle

\begin{abstract}
In this work we analyze the motion of charged particles in the background of the Kiselev geometry, which is considered here as an exact solution in the context of power-Maxwell electrodynamics. As it is well known, one can use either an electric ansatz or a magnetic one for the nonlinear electromagnetic field. We study the motion of an electrically charged particle for an electrically charged black hole and also for a magnetically charged black hole. In the second case the motion is restricted to Poincar\'e cones of various angles, as expected.
\end{abstract}

\baselineskip 1.5em

\newpage

\section{Introduction}

There is compelling evidence that the Universe is currently undergoing a period of accelerated expansion \cite{Riess} - \cite{BAO1}. What causes this mysterious expansion is still unknown and it is usually designated by the term dark energy.  As such, the study of dark energy is an important topic of research in both astrophysics and modern cosmology. Dark energy is distinguished from ordinary matter by the fact that it has negative pressure. This negative pressure leads to the accelerated expansion of the Universe by counteracting the attractive gravitational force.

The simplest explanation of dark energy was advanced by Einstein \cite{ORaifeartaigh:2017uct} in 1917 in the form of a cosmological constant, which should correspond to the energy associated with the vacuum (see the review \cite{Padmanabhan:2002ji} and references therein). To characterize the accelerated expansion one often uses an equation of state of the form $w=\frac{P}{\rho}$, where $P$ is an isotropic pressure and $\rho$ is the energy density of dark energy. Then, a cosmological constant would simply correspond to an equation of state with $w=-1$ while the energy density is constant.

There is also the possibility that the accelerated expansion is driven by a dynamical field, such as quintessence \cite{Capozziello:2005ra,Vikman:2004dc} (see also \cite{Tsujikawa:2013fta} and \cite{Wolf:2023uno}): a canonical scalar field coupled to gravity and whose potential is decreasing as the field increases. A slowly varying scalar field $\phi$ with an appropriate scalar potential $V(\phi)$ can lead to the accelerated expansion of the Universe \cite{Capozziello:2005mj}.

Twenty years ago, a spherically-symmetric static solution of Einstein equations, describing black holes surrounded by `quintessence'-like fluids was found by Kiselev \cite{Kiselev:2002dx}. It was later understood that the dark energy fluid in the Kiselev geometry is actually anisotropic \cite{Visser:2019brz} and that it is characterized by a radial pressure $p_r=-\rho_0$ and two tangential pressures $p_t$. If one defines the isotropic pressure $P=\frac{p_r+2p_t}{3}$ then the equation of state becomes $P=w\rho_0$, where $w$ is a constant parameter named as the quintessence parameter in this context. If the fluid behaves like dark energy then the quintessence parameter should have values in the interval  $-1<w<-\frac{1}{3}$. 

More recently, the Kiselev geometry has been reinterpreted in the context of nonlinear electrodynamics \cite{Dariescu:2022kof}. The study of nonlinear electrodynamics theories has a long history, as they were introduced initially in order to cure the infinite electric field and the infinite self-energy for point-like charged particles \cite{Born} and as effective classical modifications from QED \cite{Euler}. Nowadays, the nonlinear electrodynamics theories are an active area of research and they provided us with various interesting classes of black hole solutions in four and higher dimensions (for a review and more references see \cite{Breton:2007bza} and \cite{Bokulic:2021dtz}). 

In \cite{Dariescu:2022kof} the study was confined to the so-called power-Maxwell theory \cite{Hassaine:2008pw} - \cite{Hendi:2010zza}. Instead of the usual Maxwell Lagrangean $L=-F$ in this case one considers a more general theory of the form $L=-\alpha F^q$, where $q$ is a general parameter while the coupling constant $\alpha$ has to be introduced to assure a positive energy density for the nonlinear electromagnetic field. This class of theories belongs to a restricted class of nonlinear electrodynamics theories for which the Lagrangian $L=L(F)$ depends only on the first electromagnetic invariant, $F=F_{\mu\nu}F^{\mu\nu}$ and not on the second invariant $F_{\mu\nu}(\star F)^{\mu\nu}$. This greatly simplifies the field equations of the nonlinear electromagnetic field. The Kiselev solution in power-Maxwell theory \cite{Dariescu:2022kof} corresponds then to a black hole in the power-Maxwell theory, however, for the nonlinear electromagnetic field to act effectively as a dark energy fluid one has to restrict the values of the parameters accordingly. In particular, the Kiselev geometry can have two horizons, one black hole horizon and one `cosmological'-like horizon, just as it happens in the Schwarzschild - de Sitter geometry. The static patch of the Kiselev geometry is then restricted to lie in between these two horizons.

The purpose of this paper is to study the trajectories of charged massive particles moving around a Kiselev black hole in the power-Maxwell theory\footnote{Charged particles trajectories around an electrically charged Reissner-Nordstr\"{o}m black hole with quintessence have been considered in \cite{Jeong:2023hom}. In our work the `quintessence' itself is regarded as a nonlinear electromagnetic field.}. Note that this geometry can be sourced by a nonlinear electromagnetic field using either an electric ansatz (in which case the black hole is electrically charged) or a magnetic monopole ansatz (in which case the black hole carries a magnetic charge). The motion of the charged particles will obviously have different characteristics in each case. For the electrically charged case, due to the spherical symmetry of the system the charged particle motion is planar and can be studied in the equatorial plane $\theta=\frac{\pi}{2}$. In the magnetic case the motion will be confined to the Poincar\'e cones of various angles. This is in fact to be expected due to the $SO(3)$ symmetry of the system and it has been long noticed since the works of Poincar\'e \cite{Poincare},  (see \cite{Lim:2022qrt}, \cite{Lim:2021ejg} and references there).

The structure of this paper is as follows: in the next section we introduce the Kiselev geometry as a solution of the power-Maxwell theory and discuss some of its properties that will be useful later. In section $3$ we discuss the trajectories of the charged particles moving in the background of an electrically charged Kiselev black hole. In Section $4$ we address the motion of charged particles in the background of the magnetically charged Kiselev black hole. The last section is dedicated to conclusions and avenues for further work.

\section{The Kiselev black hole in the power-Maxwell theory}

The full action of the power-Maxwell theory is given by \cite{Hassaine:2008pw}, \cite{Gonzalez:2009nn}:
\beqs
I=-\frac{1}{16\pi G}\int_{\cal V}d^4x\sqrt{-g}\left(R-\alpha F^q\right)-\frac{1}{8\pi G}\int_{\partial {\cal V}}d^3x\sqrt{-\gamma}K+I_{bd},
\label{actionfull}
\eeqs
where we denoted $F=F_{\mu\nu}F^{\mu\nu}$ and $K$ is the usual Gibbons-Hawking boundary term, defined on the spacetime boundary $\partial {\cal V}$, on which the induced metric is denoted by $\gamma_{ab}$. The terms $I_{bd}$ refer to possible counterterm-like terms (for the gravitational and/or electromagnetic fields) needed to render the full action (\ref{actionfull}) finite.

The field equations derived from this action can be written in the form:
\beqs
G_{\mu\nu}=T_{\mu\nu}\\
\partial_{\mu}\left(\sqrt{-g}F^{\mu\nu}F^{q-1}\right)&=&0,
\label{eom}
\eeqs
where the stress-energy tensor of the nonlinear electromagnetic field is defined as:
\beqs
T_{\mu\nu}&=&2\alpha\bigg[qF_{\mu\rho}F_{\nu}^{~\rho}F^{q-1}-\frac{1}{4}g_{\mu\nu}F^q\bigg].
\label{tem}
\eeqs

The Kiselev black hole is described by the static four-dimensional line-element \cite{Dariescu:2022kof}
\begin{equation}
ds^2= - f (r) dt^2 + f (r)^{-1} dr^2 + r^2 \left[  d \theta ^2 + \sin^2 \theta d \varphi^2 \right]  \; ,
\label{metric}
\end{equation}
with
\begin{equation}
f(r) = 1 -\frac{2M}{r} - k r^p \, ,
\end{equation}
where the arbitrary parameter $p$ can be related to the equation of state parameter in the original Kiselev solution by
$p=-(3w+1)$.\footnote{Note that $p$ is a parameter here and should not be confused with the pressures in the equation of state.}  Then since $w\in[-1 , -\frac{1}{3})$ one obtains the range of powers $p \in (0, 2]$. It is only in this regime that the nonlinear electromagnetic field acts effectively as a dark energy source. In the power-Maxwell theory the constant $p$ is actually related to the power $q$ that appears in the power-Maxwell Lagrangian. However, unlike the usual Maxwell theory, it turns out that the power $q$ differs considerably in the electric and the magnetic cases \cite{Dariescu:2022kof}. More specifically, for the electric case one has $q=\frac{p-2}{2p}<0$, while in the magnetic case one has $q=\frac{2-p}{4}>0$. In consequence, the parameter $\alpha=(-1)^{-q}$ in the electric case, while $\alpha=1$ when using the magnetic ansatz.

In the original Kiselev geometry \cite{Kiselev:2002dx}  the positive parameter $k$  is related to the quintessence-fluid energy density $\rho_0$ as
\[
k = - \frac{\rho_0}{3w} \;,
\]
however in the power-Maxwell context the parameter $k$ will have a different interpretation, as one can see bellow.

Now, as the source of the Kiselev geometry (\ref{metric}), one may consider a nonlinear electric field given by
the electromagnetic potential component:
\begin{equation}
A_t \, = C_1 + \, C_2 r^{p+1} \; ,
\label{elem}
\end{equation}
where $C_{1,2}$ are constants of integration and moreover the constant $C_2$ is related to the electric charge $Q_e$ by \cite{Dariescu:2022kof}:
\[
C_2 \, = \, \frac{Q_e^{-\frac{p}{2}}}{2^{\frac{p+2}{4}}(p+1)} \; .
\]

In \cite{Dariescu:2022kof}, it was further shown that the positive parameter $k$ can be related to the electric charge of the nonlinear electromagnetic field
\begin{equation}
k \, = \, \frac{\left(2Q_e^2\right)^{\frac{2-p}{4}}}{p(p+1)}. 
\end{equation}

Alternatively, one may consider a magnetic monopole ansatz which also satisfies the nonlinear Maxwell equations,
with the essential component of the four-potential:
\begin{equation}
A_{\varphi} = Q_m(1-\cos\theta) \; ,
\label{magem}
\end{equation}
where $Q_m$ is the magnetic charge \cite{Dariescu:2022kof}. In this particular case we have:
\begin{equation}
k \, = \,  \frac{\left(2Q_m^2\right)^{\frac{2-p}{4}}}{2(p+1)} \; .
\end{equation}

Depending on the values taken by $M$, $p$ and $k$ one can have at most two horizons, namely the black hole horizon $r_b$ and an effective cosmological horizon, located at $r_c > r_b$. As the black hole mass parameter, $M$, is increasing, the black hole horizon $r_b$ increases while the cosmological horizon $r_c$ shrinks. For fixed values of $Q$ and $p$, there is a maximum value of the mass parameter $M$:\footnote{Note that there are some typos in the corresponding formula in \cite{Dariescu:2022kof}.}
\begin{equation}
M_{max}  \, = \, \frac{1}{2}p(p+1)^{-\frac{p+1}{p}}k^{-\frac{1}{p}} \; ,
\end{equation}
beyond which the spacetime geometry is singular. For this value of the mass the geometry becomes extremal, with both the black hole horizon $r_b$ and the cosmological horizon $r_c$ being equal.

\section{Charged particles moving in the background of the electrically charged Kiselev black hole}

In Kiselev geometry, the timelike trajectories of an uncharged particle moving around the black hole described by the
metric (\ref{metric}), can be obtained starting with the Lagrangean ${\cal L}=\frac{1}{2}g_{\mu\nu}\dot{x}^{\mu}\dot{x}^{\nu}$, where $\dot{x}^{\mu}=\frac{dx^{\mu}}{d\tau}$ while the proper time $\tau$ is defined using:
\begin{equation}
 -d \tau^2=- f(r) dt^2+\frac{dr^2}{f(r)} + r^2 (  d \theta^2 + \sin^2 \theta d \varphi^2 )  \; .
\label{tau}
\end{equation}

Due to the spherical symmetry the motion is planar and it can be confined to the equatorial plane $\theta=\frac{\pi}{2}$. Since $t$ and $\varphi$ are cyclic coordinates they lead to two constants of motion: the energy and the angular momentum.
If one replaces the conserved energy and angular momentum per unit mass, $E=f(r) \dot{t}$ and $L =  r^2 \dot { \varphi }$ in (\ref{tau}) one finds, for $\theta = \pi /2$, the relation
\begin{equation}
\dot{r}^2 = E^2  - f(r) \left[ 1 + \frac{L^2}{r^2} \right] \; .
\label{e1}
\end{equation}
which leads to the effective potential for the uncharged particles: 
\begin{equation}
V_{eff} =f(r)\left[ 1 + \frac{L^2}{r^2} \right] = \left[ 1 -\frac{2M}{r} - kr^p \right] \left[ 1 + \frac{L^2}{r^2} \right] .
\label{poteff}
\end{equation}

If one considers now a charged particle moving in the electric field generated by the electromagnetic potential (\ref{elem}) the motion will be described by the Lagrangean:
\beqs
{\cal L}=\frac{1}{2}g_{\mu\nu}\dot{x}^{\mu}\dot{x}^{\nu}+\varepsilon A_{\mu}\dot{x}^{\mu},
\eeqs
where $\varepsilon=e/ m$ is the specific charge of the test particle with charge $e$ and mass $m$. Since $t$ and $\varphi$ are still cyclical coordinates they still lead to conserved quantities. However, while the expression for the angular momentum remains the same, the expression of the energy must be replaced by
\[
E \, = \, f(r)\dot{t}+ \varepsilon A_t.
\]

Thus, the equation (\ref{e1}) is modified in the form:
\begin{equation}
\dot{r}^2 \, = \, \left(E- \varepsilon A_t \right)^2-f(r)\left(1+\frac{L^2}{r^2}\right)=(E-V_+)(E-V_-),
\end{equation}
where we defined\footnote{For the corresponding analysis in the Reissner-Nordstr\"{o}m case see \cite{Pugliese:2011py} and \cite{Grunau:2010gd}.}:
\begin{equation}
V_{\pm} \, = \, \varepsilon A_t \pm\sqrt{f(r)\left(1+\frac{L^2}{r^2}\right)}.
\label{vpcharge}
\end{equation}

Note that  $V_{\pm}$ correspond to those values of the energy per unit mass $E$ that make $r$ into a turning point, where the value of the kinetic energy of the test particle vanishes $\dot{r}^2=0$. At the horizons, the effective potentials $V_+$ and $V_-$ become equal to $V(r_{\pm}) = \varepsilon \left[ C_1 + C_2 r_{\pm}^{p+1} \right]$. 

While the $C_2>0$ constant is related to the black hole electric charge $Q_e>0$, the value of the $C_1$ constant is not fixed \textit{a priori}. While during the studies of the motion of charged test particles in the usual Reissner-Nordstr\"{o}m geometry  \cite{Pugliese:2011py}, \cite{Grunau:2010gd}, \cite{Pugliese:2011py} its value is set to zero, in our case, we have chosen to use it to set the value of the effective potential (\ref{vpcharge}) to be zero on the black hole horizon $r=r_-$, that is $C_1=-C_2r_{-}^{p+1}$. With this choice one obtains $A_t=C_1+C_2r^{p+1}=C_2(r^{p+1}-r_-^{p+1})>0$ since $r>r_-$ outside the black hole horizon. Therefore, the sign of the first term $\varepsilon A_t$ in (\ref{vpcharge}) is controlled by the sign of $\varepsilon$.

Note that $V_+(\varepsilon, L, r)\geq V_-(\varepsilon, L, r)$ and one also has $V_+(\varepsilon, L, r)=-V_-(-\varepsilon, L, r)$. Moreover, for $\varepsilon=0$ the potential $V_+$ reduces to the effective potential for uncharged test particles (\ref{poteff}).

In the followings, for future-directed orbits for the charged particles, we shall use the effective potential $V_+$ to study the bound motion of the charged timelike particles, for different ranges of the model's parameters.

\subsection{Circular orbits}

In order to have a circular motion, in the region between the two horizons, one has to impose
the conditions
\begin{equation}
V_+ = E \; , \quad  \frac{dV_+}{dr} = 0 \; ,
\label{circ} 
\end{equation}
Following the approach developed in \cite{Pugliese:2011py}, one could solve the second equation in the relation above to find the angular momentum $L$ of the charged particle on the circular orbit of radius $r = R_c$. The corresponding energy $E$ can be found by substituting this value of the angular momentum in the first equation in (\ref{circ}). However, unlike the case corresponding to the uncharged particle which has been discussed in \cite{Dariescu:2022kof}, the situation here is much more complicated, due to the additional term $\frac{e}{m}A_t$.

For example, for $p=1$, the potential takes the form
\begin{equation}
V_+ =  \varepsilon \left[C_1 + C_2 r^2 \right] + \sqrt{ f(r)  \left(1+\frac{L^2}{r^2}\right)} \; ,
\label{VPel}
\end{equation}
where the constants $C_2$ and $k$ are related to the electric charge by $C_2 = 2^{-7/4} / \sqrt{Q_e}$ and $k= 2^{-3/4} \sqrt{Q_e}$.
The metric function becomes
\begin{equation}
f(r) =  1 - \frac{2M}{r} - kr \, ,
\label{pequal1}
\end{equation}
pointing out the existence of the two horizons:
\begin{equation}
r_{\pm} = \frac{1\pm \sqrt{1-8kM}}{2k} \; ,
\end{equation}
for $8kM<1$.

The relation $V_+^{\prime} =0$ is leading to the following equation for the conserved momentum per unit mass
\begin{eqnarray}
& &
\Sigma^2 L^4 - 2 r^2 \left[ \left( 2M-kr^2 \right) \Sigma+ \varepsilon^2 \frac{r^6 f}{4k^2} \right] L^2
+ r^4 \left[ ( 2M-kr^2)^2- \varepsilon^2 \frac{r^6 f}{2k^2} \right] =0
\label{circLel}
\end{eqnarray}
with the notation
\begin{equation}
\Sigma = -kr^2+2r-6M. 
\end{equation}
The solutions of (\ref{circLel}) are:
\begin{eqnarray}
L^2 & = & \frac{r^2(2M-kr^2)}{\Sigma} + \frac{\varepsilon^2 r^8 f}{4 k^2 \Sigma^2} \left[ 1 \pm \sqrt{1+ \frac{16k^2 \Sigma}{\varepsilon^2 r^5}} \right]
\end{eqnarray}
and one has to impose that the energy and
angular momentum are both real and finite, in order for the circular motion of the test particle to be possible.
To first order in $k^2 /\varepsilon^2$, one may use for $L^2$ the positive expression: 
\[
L^2 \approx \frac{r^2}{\Sigma} \left[ 2r -2M -3 k r^2 \right] + \frac{\varepsilon^2r^8 f}{2 k^2 \Sigma^2}.
\]

The region where one has a circular motion has as boundaries the solutions of $\Sigma =0$. We consider the root which is in between the two horizons, i.e.
\begin{equation}
r_* = \frac{1-\sqrt{1-6kM}}{k}
\end{equation}
and, together with the condition $f(r)>0$, one finds the allowed range of the circular motion radius:
\begin{equation}
\frac{1-\sqrt{1-6kM}}{k} < R_c < \frac{1+\sqrt{1-8kM}}{2k}  \;  .
\end{equation}

For neutral particles ($\varepsilon =0$), the angular momentum has the simple form
\begin{equation}
L^2 = \frac{r^2(2M-kr^2)}{\Sigma},
\end{equation}
which is positive for $r< r_0 = \sqrt{2M/k}$, where the special (maximum) radius $r_0$ is inbetween the two horizons.
The neutral particle with zero momentum is moving on the circular orbit with radius $r=r_0$, with the energy
\[
E_0 = \sqrt{1- \sqrt{8kM}} .
\]

\begin{figure}
  \centering
  \includegraphics[width=0.45\textwidth]{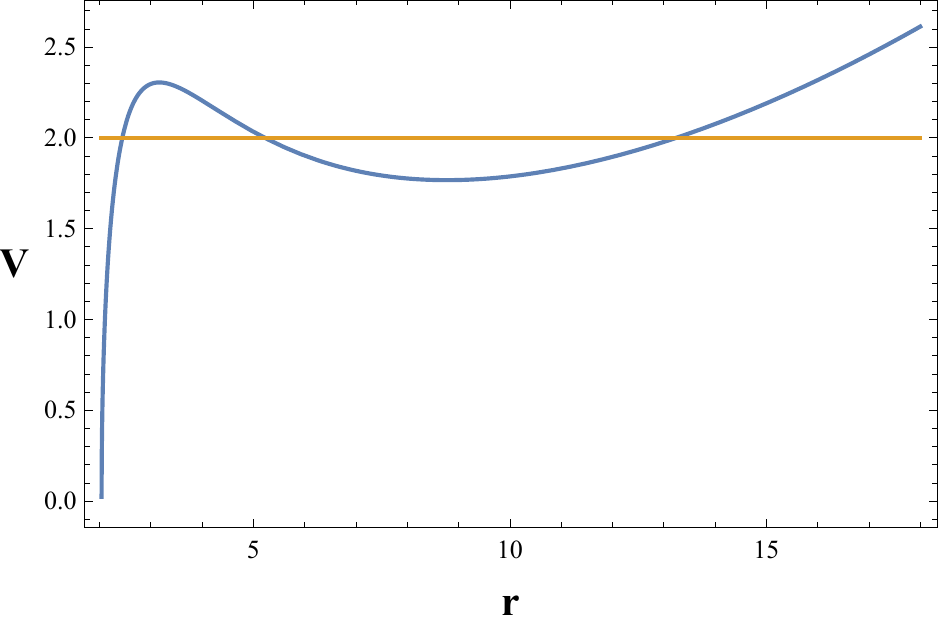}
  \caption{ The blue plot corresponds to the effective potential (\ref{VPel}) with the metric function (\ref{pequal1}). The numerical values of the parameters are: $M=1$, $\varepsilon =0.7$, $L=12$, $k=0.01$, $C_2 =0.005$ and $C_1 = -0.02$. The horizontal line representes the particle's energy.}
  \label{VFig}
\end{figure}

\begin{figure}
  \centering
  \includegraphics[width=0.45\textwidth]{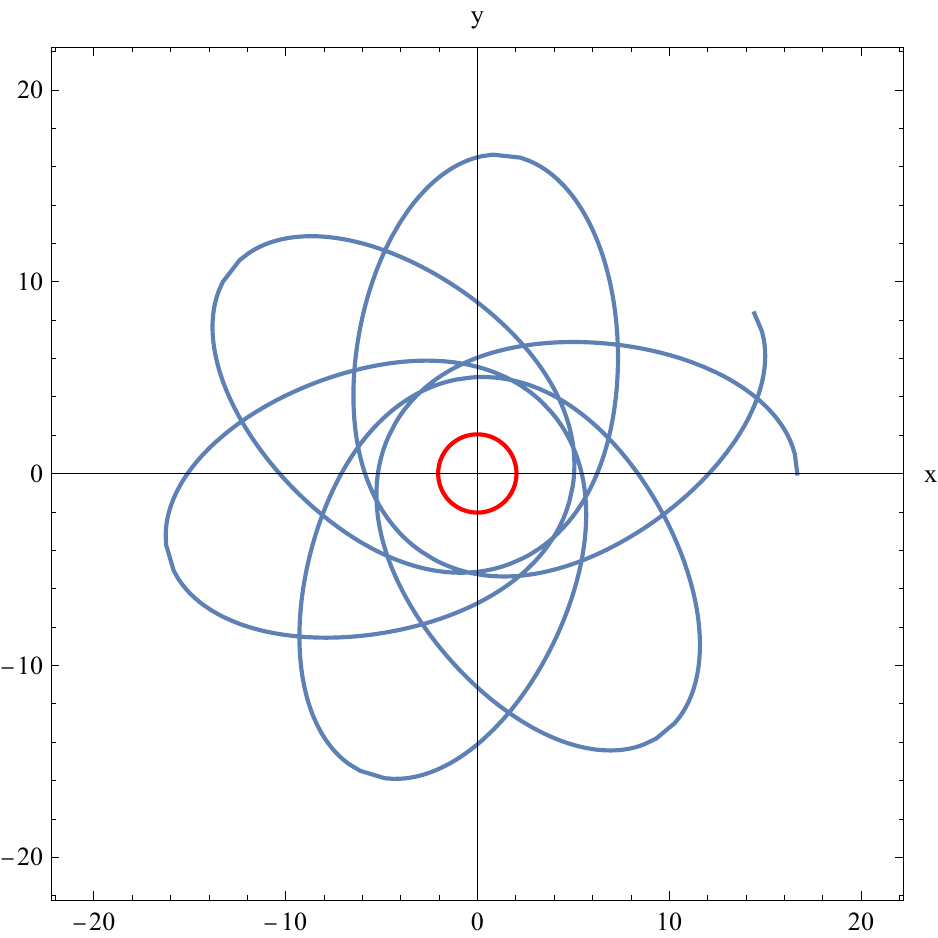}  
  \caption{Parametric plot of the bounded trajectory of the charged particle trapped by the potential (\ref{VPel}). The red circle represents the black hole horizon. The numerical values of the parameters are the same as for the Figure 1.} 
  \label{Bound}
\end{figure}

In Figure \ref{VFig}, the blue plot represents the effective potential (\ref{VPel}) while the particle's energy is represented by the horizontal line. For the imposed numerical values: $M=1$, $\varepsilon =0.7$, $L=12$ and $k=0.01$, the other parameters are: $C_2= k/2 =0.005$ and $C_1 = -C_2 r_-^2 =-0.02$ and the two horizons are situated at: $r_- = 2.04$ and $r_+ = 97.95$.
This potential leads to three turning points given by the intersection of the horizontal line with the potential and they are denoted here by $r_1<r_2<r_3$. For $E=V_{min}$, the particle is following a stable circular trajectory, while for $E=V_{max}$, the particle has an unstable circular motion. The particle with the $V_{min} < E < V_{max}$ which starts its journey between the two turning points $r_2$ and $r_3$ has the bounded periodic motion represented in the Figure \ref{Bound}. The trajectory is between the two horizons, the black hole horizon and the cosmological-like horizon. 

Note that the main characteristics of the effective potential in Figure \ref{VFig} and the bounded motion in Figure \ref{Bound} do not change if the test particle is extremally charged $\varepsilon=1$, or if it is overcharged $\varepsilon>1$.

\section{Charged particles in the background of the Kiselev geometry sourced by a magnetic monopole}

In the case of a charged particle moving in a magnetic field sourced by the potential component $A_{\varphi}$ defined in (\ref{magem}), we start with the Lagrangean: 
\begin{equation}
{\cal L} = \frac{1}{2} \left[ - f(r) \dot{t}^2 + \frac{1}{f(r)} \dot{r}^2 + r^2 ( \dot{\theta}^2+ \sin^2 \theta \dot{\varphi}^2 ) \right] + \varepsilon Q_m \dot{\varphi} (1- \cos \theta ),
\label{Lanmag}
\end{equation}
together with the normalization condition coming from $g_{\mu\nu}\dot{x}^{\mu}\dot{x}^{\nu}=-1$ for timelike particles: 
\begin{equation}
\frac{1}{f(r)} \dot{r}^2 + r^2 \dot{ \theta}^2 + r^2 \sin^2 \theta \dot{\varphi}^2 - f(r) \dot{t}^2 = -1 \; ,
\label{solder}
\end{equation}
where $dot$ means the derivatives with respect to the proper time $\tau$.
With dimensionless quantities (as they were rescaled against $r_S =2M$), one can work out the corresponding Euler--Lagrange equations. Thus, for conserved energy and angular momentum, one finds the relations:
\begin{equation}
\frac{d t}{d \tau}  = \frac{E}{f}  \Rightarrow  \frac{d t}{d \gamma}  = r^2\frac{E}{f}
\end{equation}
and
\begin{eqnarray}
\frac{d \varphi}{d \tau} = \frac{1}{r^2 \sin^2 \theta} \left[ L - \varepsilon Q_m (1 -  \cos \theta ) \right] \Rightarrow \frac{d \varphi}{d \gamma} = \frac{1}{\sin^2 \theta} \left[ L - \varepsilon Q_m (1 -  \cos \theta ) \right],
\label{varpp}
\end{eqnarray}
where we have used the Mino time $\gamma$  \cite{Mino:2003yg}, introduced here by the relation $d \tau = r^2 d \gamma$. Note that the Mino time is not an affine parameter.

Using the Carter constant defined as \cite{Baines:2021qaw}:
\beqs
K &= &r^4 \dot{\theta}^2 + r^4\sin^2\theta \dot{\varphi}^2 = r^4 \dot{\theta}^2 + \frac{ \left[  L - \varepsilon Q_m (1 -  \cos \theta ) \right]^2}{\sin^2 \theta}, 
\label{Carterc}
\eeqs
one obtains the relation:
\begin{equation}
\left( \frac{d \theta}{d \tau} \right)^2  = \frac{1}{r^4} \left[  \kappa - \frac{\left[ L - \varepsilon Q_m (1 - \cos \theta ) \right]^2}{\sin^2 \theta} \right]
\Rightarrow
\left( \frac{d \theta}{d \gamma} \right)^2  = \kappa - \frac{\left[ L - \varepsilon Q_m (1 - \cos \theta ) \right]^2}{\sin^2 \theta}
\label{thetap}
\end{equation}
where $\kappa = K/(4M^2)$ is the rescaled Carter constant.

By replacing the above results in the normalization condition (\ref{solder}), one obtains the equation describing the radial motion,
\begin{equation}
\left( \frac{d r}{d \gamma} \right)^2  = r^4 \left[ E^2 - f \left( 1 + \frac{\kappa}{r^2} \right)  \right].
\label{rpun}
\end{equation}

\subsection{A note on the Poincar\'e cones}

As we have mentioned in the Introduction, the motion of an electrically charged particle in the field of magnetic monopole should be confined on the so-called Poincar\'e cones. Generically, these cones arise from the $SO(3)$ symmetry of the system (even if the Lagrangean ${\cal L}$ does not exhibit this symmetry), which leads to various conserved quantities which can be used to show that the trajectories lie on cones. 

To this end one should note that the electrically charged particle has an angular momentum $\vec{S}$ with constant magnitude proportional to $-\varepsilon Q_m$, which is directed along the radial direction. On the other hand, the orbital angular momentum $\vec{L}$ is orthogonal to the radial direction and its magnitude $|\vec{L}|$ is also conserved (being related to the Carter constant, as we shall see bellow), while its direction changes along the trajectory. However, the total angular momentum $\vec{J}=\vec{S}+\vec{L}$ is conserved along the trajectory. This means that the trajectory of the charged particle in the field of a magnetic monopole will be confined to a cone around the direction of $\vec{J}$ defined by the constant angle $\cos\alpha=\frac{|\vec{S}|}{|\vec{J}|}$.

To see how this is working in practice, let us notice that the Kiselev geometry (\ref{metric}) admits four Killing vectors: one timelike $\xi^{\mu}_{(t)}=(1, 0, 0, 0)$, and three spacelike Killing vectors adapted to the spherical symmetry of the system:
\beqs
\xi_{(1)}^{\mu}&=&(0, 0, -\sin\varphi, -\cos\varphi \cot\theta), \nonumber\\
\xi_{(2)}^{\mu}&=&(0, 0, \cos\varphi, -\sin\varphi \cot\theta),\nonumber\\
\xi_{(3)}^{\mu}&=&(0, 0, 0, 1).
\eeqs
If the motion of the particle were geodesic then these Killing vectors will all lead to conserved quantities along the motion, however, this is not the case here due to the electromagnetic Lorentz forces. In fact, these Killing vectors can be used to define the components of the orbital angular momentum along the charged particle's trajectory: 
\beqs
L_{x}&=&-r^2\sin\varphi~\dot{\theta}-r^2\sin\theta\cos\theta\cos\varphi~\dot{\varphi},\nonumber\\
L_{y}&=&r^2\cos\varphi~\dot{\theta}-r^2\sin\theta\cos\theta\sin\varphi~\dot{\varphi},\nonumber\\
L_{z}&=&r^2\sin^2\theta~\dot{\varphi}.
\eeqs
If one defines the components of the angular momentum $\vec{S}=-\varepsilon Q_m \hat{e}_r$ as:
\beqs
S_x&=&-\varepsilon Q_m\sin\theta\cos\varphi,\nonumber\\
S_y&=&-\varepsilon Q_m\sin\theta\sin\varphi,\nonumber\\
S_z&=&-\varepsilon Q_m\cos\theta,
\eeqs
then the total angular momentum of the system is now defined as:
\beqs
\vec{J}&=&\vec{S}+\vec{L}
\eeqs
and it can be checked that it is a constant throughout the charged particle motion: $\frac{d\vec{J}}{d\tau}=\vec{0}$. Note that the constant of motion $L$ derived using the Euler-Lagrange equation in (\ref{varpp}) can be related to the $z$-component $J_z$ of the total angular momentum $\vec{J}$ as $J_z=L-\varepsilon Q_m$ hence $J_z$ is a constant of motion. To check that the other components satisfy $\frac{dJ_x}{d\tau}=\frac{dJ_y}{d\tau}=0$ one has to use the following Euler-Lagrange equations for the coordinates $\theta$ and $\varphi$, namely:
\beqs
r^2\ddot{\theta}+2r\dot{r}\dot{\theta}-r^2\sin\theta\cos\theta\dot{\varphi}^2-\varepsilon Q_m\sin\theta\dot{\varphi}&=&0,
\label{Eulertheta}
\eeqs
respectively
\beqs
r^2\sin\theta\ddot{\varphi}+2r\sin\theta\dot{r}\dot{\varphi}+2r^2\cos\theta\dot{\theta}\dot{\varphi}+\varepsilon Q_m\dot{\theta}&=&0.
\label{Eulervarphi}
\eeqs
Squaring the total angular momentum one finds:
\beqs 
J^2&=&|\vec{L}|^2+(\varepsilon Q_m)^2,
\eeqs
which means that the magnitude of the orbital angular momentum is also conserved $|\vec{L}|=const.$ The angle between the direction of $\vec{J}$ and the radial direction $\hat{e}_r$ is constant:
\beqs
\cos\alpha&=&-\frac{\varepsilon Q_m}{J},
\eeqs
which means that the charged particle trajectory is confined on a Poincar\'e cone around the direction of $\vec{J}$.

Note that even if the components of the orbital angular momentum are not conserved along the trajectory of the charged particle, if one considers the magnitude of the angular momentum vector $\vec{L}$:
\beqs
|\vec{L}|^2&=&L_x^2+L_y^2+L_z^2=r^4\dot{\theta}^2+r^4\sin^2\theta~\dot{\varphi}^2
\eeqs
one can notice that its magnitude equals precisely the value of the Carter constant (\ref{Carterc}). This is no accident, since for spherically symmetric geometries of the form (\ref{metric}) there exists one quadratic Killing vector (besides the trivial Killing tensor given by the metric $K_{\mu\nu}=g_{\mu\nu}$) of the form:
\beqs
K^{\mu\nu}&=&\xi_{(1)}^{\mu}\xi_{(1)}^{\nu}+\xi_{(2)}^{\mu}\xi_{(2)}^{\nu}+\xi_{(3)}^{\mu}\xi_{(3)}^{\nu}
\eeqs
and it corresponds directly to the Carter constant $K$ in (\ref{Carterc}). As a consequence, one can write $J=\sqrt{K+(\varepsilon Q_m)^2}$. Finally, the angle between the cone axis and the $Oz$ axis is:
\beqs
\cos \psi&=&\frac{J_z}{J}=\frac{L-\varepsilon Q_m}{\sqrt{K+(\varepsilon Q_m)^2}}.
\eeqs

\subsection{The $\theta$ motion and the allowed range of $L-$values}

By inspecting the relation (\ref{thetap}), one can notice that the right hand side of the equation should be a positive quantity. For $\kappa >0$, one has
the following inequality:
\begin{equation}
\left[ \kappa + ( \varepsilon Q_m)^2 \right] \cos^2 \theta + 2 \varepsilon Q_m (L- \varepsilon Q_m) \cos \theta + (L - \varepsilon Q_m)^2 - \kappa \leq 0.
\label{thetaplus}
\end{equation}
For $\theta =0$, the relation (\ref{thetaplus}) is satisfied only for $L=0$, while for $\theta = \pi$, the corresponding angular momentum is $L=-2 \varepsilon Q_m$.

One has to impose the discriminant $\Delta \geq 0$ and $\cos \theta$ in between the two real roots:
\begin{equation}
\cos \theta_{1,2} = \frac{- \varepsilon Q_m(L-\varepsilon Q_m) \pm \sqrt{\Delta}}{\kappa+(\varepsilon Q_m)^2}
\label{thetaturn}
\end{equation}
with
\begin{equation}
\Delta = \kappa \left[ \kappa - L (L- 2 \varepsilon Q_m) \right] 
\label{Deltath}
\end{equation}

The analytical solution of the differential equation (\ref{thetap}) can be easily obtained. With the change of function $\cos \theta =x$ and the values (\ref{thetaturn}) denoted by
$\cos \theta_1 \equiv a$ and $\cos \theta_2 \equiv b$, the relation (\ref{thetap}) can be written as
\[
- \frac{dx}{\sqrt{-(x-a)(x-b)}} = \sqrt{\kappa + ( \varepsilon Q_m)^2} \; d \gamma
\]
and leads, by integration, to
\begin{eqnarray}
\theta (\gamma ) & = & \arccos \left[  \frac{a+b}{2} + \frac{a-b}{2} 
\cos \left[ \sqrt{\kappa + (\varepsilon Q_m)^2} (\gamma - \gamma_0 )  \right]  \right] \nonumber \\*
& = &
\arccos \left[  \frac{- \varepsilon Q_m(L-\varepsilon Q_m)}{\kappa+(\varepsilon Q_m)^2} - \frac{\sqrt{\Delta}}{\kappa+(\varepsilon Q_m)^2}  \cos \left[ \sqrt{\kappa + (\varepsilon Q_m)^2} (\gamma - \gamma_0 )  \right] \right]
\label{thetag}
\end{eqnarray}
By integrating from $\cos \theta_1 = a$ to $\cos \theta_2 =b$, we obtain the periodicity: 
\[
\Delta \gamma = \frac{\pi}{ \sqrt{\kappa + ( \varepsilon Q_m)^2}}
\]

\begin{figure}
  \centering
  \includegraphics[width=0.45\textwidth]{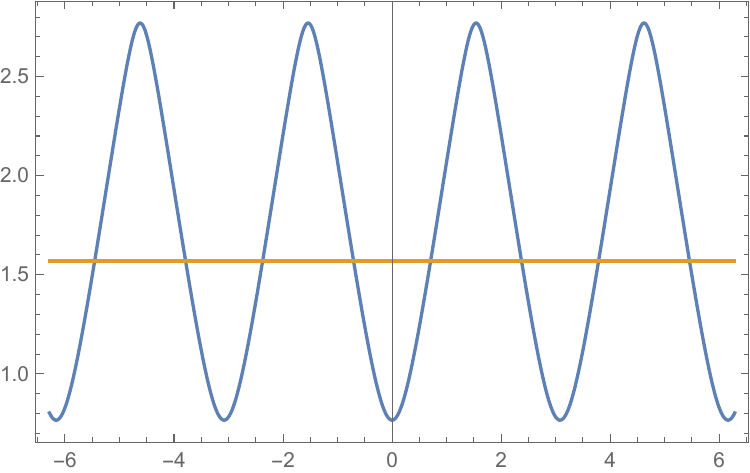}
  \caption{The  temporal evolution of $\theta ( \gamma )$ given in (\ref{thetag}).
The numerical values of the dimensionless parameters are: $M=1$, $\varepsilon Q_m = 0.4$, $\kappa =4$, $L=1.5$. The horizontal line corresponds to $\theta=\frac{\pi}{2}$.}
\label{fig3}
\end{figure}

In Figure \ref{fig3}, we have plotted the temporal evolution of the $\theta-$coordinate given in (\ref{thetag}). For specific values of the parameters, this is oscillating between $\theta_1 =0.755$ and $\theta_2 = 2.77$ and is periodically passing the $\theta = \pi/2$ line.

Keeping the same notations, one can derive from the equations (\ref{varpp}) and (\ref{thetap}), the following solution of the $\varphi-$equation:
\begin{eqnarray}
\varphi (\gamma ) & = & - \frac{1}{\sqrt{\kappa +   (\varepsilon Q_m)^2}} \int \frac{\left[ L -  \varepsilon Q_m (1-x) \right] dx}{(1-x^2) \sqrt{-(x-a)(x-b)}}
\nonumber \\*
& = &
\varphi_0 + 
\arctan \left[ \sqrt{\frac{1+b}{1+a}} \sqrt{\frac{x-a}{b-x}} \right]
- \arctan \left[ \sqrt{\frac{1-b}{1-a}} \sqrt{\frac{x-a}{b-x}} \right].
\end{eqnarray}
Note that for $x=a$ or $x=b$ one has $\varphi=\varphi_0=constant$, which means that the motion is purely radial on a cone.
 
Going back to the relations (\ref{thetaturn}) and (\ref{Deltath}), one can notice that the physical conditions $\Delta \geq 0$ and $\cos \theta_{1,2} \in [-1 , 1]$ are leading to the range of the angular momentum:
\begin{equation}
L \in \left[ \varepsilon Q_m - \sqrt{\kappa+(\varepsilon Q_m)^2} ,  \varepsilon Q_m + \sqrt{\kappa+(\varepsilon Q_m )^2}  \right].
\label{inequalityL}
\end{equation}
Depending on the signs of $\varepsilon Q_m$ and the one in front of $\sqrt{\Delta}$ in (\ref{thetaturn}), the particle can cross the
equatorial plane. The values
\[
L_{\pm} =   \varepsilon Q_m \pm \sqrt{\kappa +(\varepsilon Q_m)^2}
\]
correspond to $\Delta =0$.
For $L=L_+$ which is positive for both $\varepsilon Q_m >0$ and $\varepsilon Q_m <0$, the only possible value of $\theta$ is given by
\[
\cos \theta_+ = - \, \frac{\varepsilon Q_m}{\sqrt{\kappa+( \varepsilon Q_m)^2}} .
\] 
Depending on the sign of $\varepsilon Q_m$, the angle $\theta_+$ can be either $\theta_+ \in \left( 0 , \frac{\pi}{2} \right)$ or
$\theta_+ \in \left( \frac{\pi}{2} , \pi \right)$.

For $L=L_-$, which is a negative quantity, the value of $\theta$ is
\[
\cos \theta_- = \frac{\varepsilon Q_m}{\sqrt{\kappa+( \varepsilon Q_m)^2}}. 
\] 

The $\theta$ motion can then be classified according to the range of the angular momentum $L$.
Let us consider $\varepsilon Q_m >0$, but similar conclusions can be drawn for $\varepsilon Q_m <0$.
For the negative range
\begin{equation}
\varepsilon Q_m - \sqrt{\kappa + ( \varepsilon Q_m)^2} < L < \varepsilon Q_m - \sqrt{\kappa}
\end{equation}
the particle is above the equatorial plane, i.e. $\theta \in \left( 0 , \frac{\pi}{2} \right)$.
For
\begin{equation}
 \varepsilon Q_m - \sqrt{\kappa} < L <   \varepsilon Q_m + \sqrt{\kappa} 
\end{equation}
the values $\cos \theta_{1,2}$ in (\ref{thetaturn}) have opposite signs and the particle crosses the equatorial plane.
Finally, for 
\begin{equation}
 \varepsilon Q_m + \sqrt{\kappa} < L <    \varepsilon Q_m + \sqrt{\kappa + ( \varepsilon Q_m)^2}
 \label{equa} 
\end{equation}
the particle is bellow the equatorial plane, i.e. $\theta \in \left( \frac{\pi}{2} , \pi \right)$. The particle with $L =   \varepsilon Q_m + \sqrt{\kappa}$ moves on the equatorial plane.

For $L$ outside the range (\ref{inequalityL}), i.e. $L <  \varepsilon Q_m - \sqrt{\kappa + ( \varepsilon Q_m)^2} $ or $L >  \varepsilon Q_m + \sqrt{\kappa + ( \varepsilon Q_m)^2} $, the inequality (\ref{thetaplus}) is not satisfied and the motion is not possible.

\subsection{The radial motion and circular trajectories}

Turning now our attention to the radial motion, one has to impose that the right hand side of the relation (\ref{rpun}) is positive, i.e.
\[
kr^4 + (E^2 -1) r^3 + (1+k \kappa) r^2 - \kappa r +\kappa \geq 0,
\]
where everything is expressed in units of $2M$.

The regions for which the above condition is satisfied are bounded
by the zeros of the fourth degree polynomial. The number and the nature of zeros are depending on the
particle's energy and on the values of parameters $k$ and $\kappa$.

One may notice that for $p=1$ the relation (\ref{rpun}) is leading to the effective potential:
\begin{equation}
V = \left( 1- \frac{1}{r} - k r \right) \left( 1 + \frac{\kappa}{r^2} \right),
\label{poteff2}
\end{equation}
where we used rescaled quantities.
This is vanishing on the two horizons present in the Kiselev geometry:
\[
r_{\pm} = \frac{1 \pm \sqrt{1-4k}}{2k}
\]
and is positive in-between. For $k \ll 1$, the values of $r_{\pm}$ can be approximated to
$r_- \approx 1$ and $r_+ \approx 1/k$.

Depending on the values of $k$ and $\kappa$, the effective potential allows attracting orbits, escape orbits, bound orbits, stable and unstable circular orbits.
These are the same as for the uncharged test particles with $L^2 =\kappa$ and have been categorized in \cite{Fathi:2022pqv}, \cite{Wang:2023otn}.

There is at least a maximum of the potential which corresponds to an unstable circular orbit.
The conditions
\[
V = E^2 \; , \quad V_{eff}^{\prime} = 0
\]
lead to a system of two equations of fourth order:
\begin{eqnarray}
& &
kr^4 + (E^2 -1) r^3 + (1+k \kappa) r^2 - \kappa r + \kappa =0, \nonumber \\*
& &
kr^4 -(1+k \kappa ) r^2 +2 \kappa r - 3 \kappa = 0.
\end{eqnarray}
With
\[
\kappa = \frac{R^2(1-kR^2)}{2R - 3 -kR^2}
\]
one may write the expression of the energy of particle on the inclined circular orbit of radius $R$ as being
\begin{eqnarray}
E^2 & =& \frac{2(R-1-kR^2)^2}{R(2R - 3 -kR^2)}  = \frac{2Rf(R)^2}{\Sigma (R) }.
\end{eqnarray}
By imposing that $E^2$ and $\kappa$ are positive quantities, one may find the range of the circular radius $R$ as being
\[
r_- < \frac{1-\sqrt{1- 3k}}{k} < R < \sqrt{\frac{1}{k}} <  r_+ .
\]

Thus, the physical range for the circular orbit is depending on the parameter $k$ and it shrinks as $k$ increases.
Also, the condition $V^{\prime \prime} (R) \geq 0$, corresponding to a stable circular orbit, leads to a relation between $k$ and $\kappa$, i.e.
\[
\kappa \geq \frac{2}{3-2k}.
\]

In order to plot the trajectory of a bound particle whose orbital momentum is in the allowed range, one has to use the system of Euler--Lagrange equations derived from the Lagrangean (\ref{Lanmag}), i.e.
\begin{eqnarray}
&&
\ddot{r} = \frac{\dot{r}^2 f^{\prime}}{2f} - \frac{f^{\prime}E^2}{2f} + r f \dot{\theta}^2 + \frac{f \left[ L - \varepsilon Q_m (1- \cos \theta) \right]^2}{r^3 \sin^2 \theta},  \nonumber \\*
& & \ddot{\theta} = - \frac{2 \dot{r} \dot{\theta}}{r} + \frac{\cos \theta \left[ L - \varepsilon Q_m (1- \cos \theta) \right]^2}{r^4 \sin^3 \theta} + \frac{\varepsilon Q_m \left[ L - \varepsilon Q_m (1- \cos \theta) \right]}{r^4 \sin \theta},
\end{eqnarray}
where $\dot{()}$ and $()^{\prime}$ are the derivatives with respect to $\tau$ and $r$, respectively.

\begin{figure}
  \centering
  \includegraphics[width=0.5\textwidth]{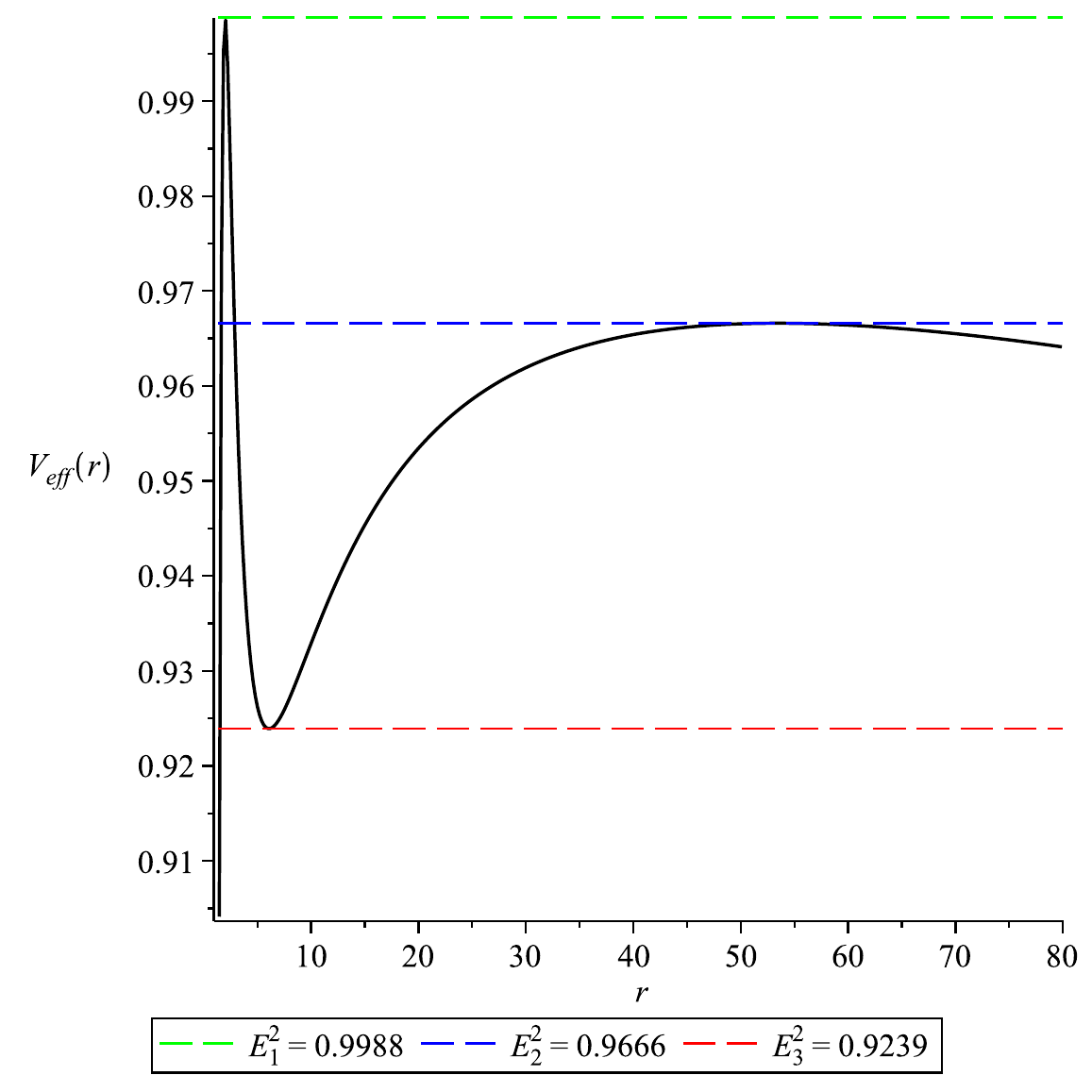}
  \caption{The effective potential (\ref{poteff2}) for $k=4$ and $\kappa=0.0003$. The horizontal lines correspond to various energies of the charged particle.}
\label{fig4}
\end{figure}

As an example, let us consider the potential (\ref{poteff2}) represented in Figure \ref{fig4},  for values of $r/(2M)$ in the range which allows the particle's bound orbits. There are two unstable circular orbits corresponding to the two maxima of the potential and one stable minimum.

\begin{figure}[ht!]
\begin{center} 
\includegraphics[width=0.43\textwidth, angle =0 ]{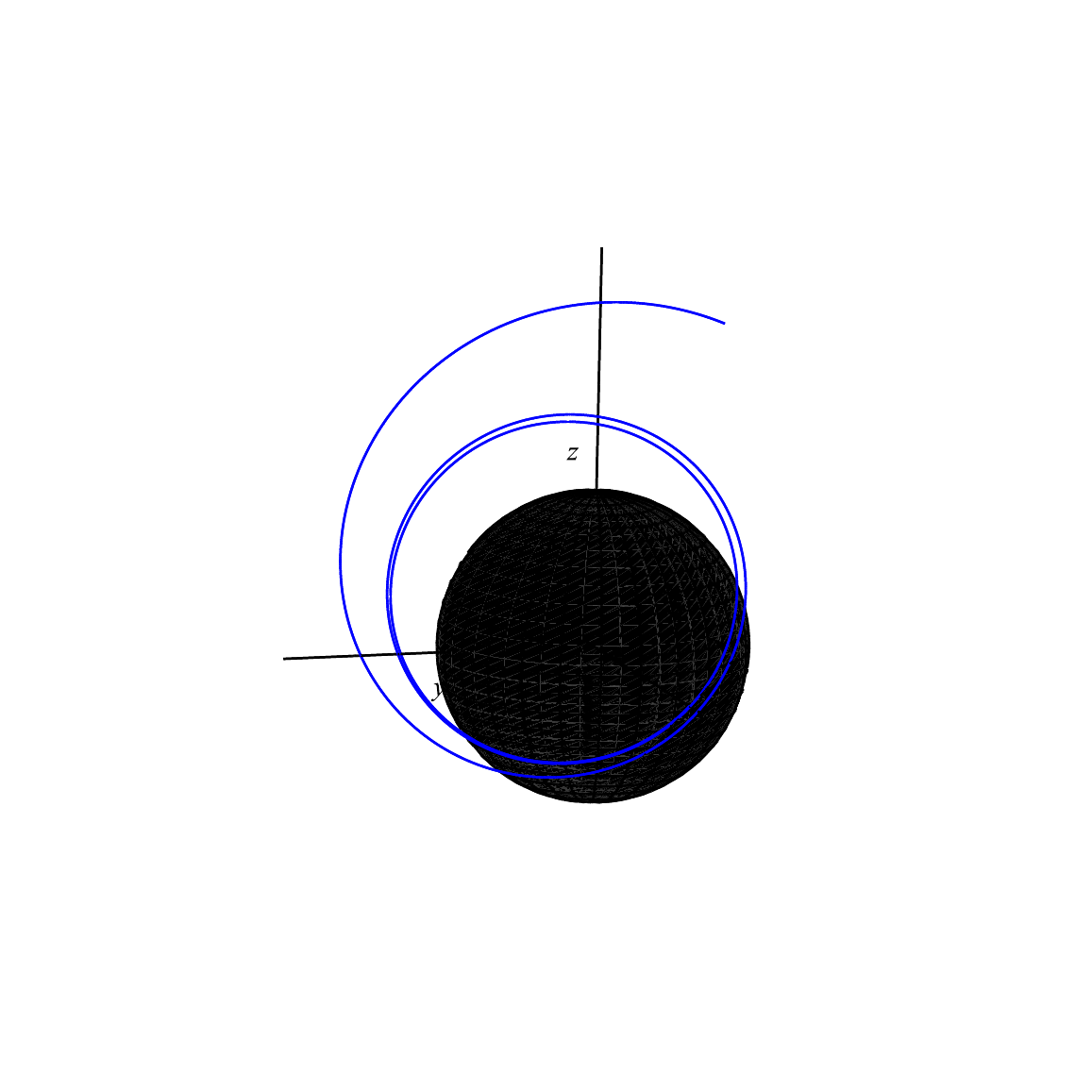}
\includegraphics[width=0.43\textwidth, angle =0 ]{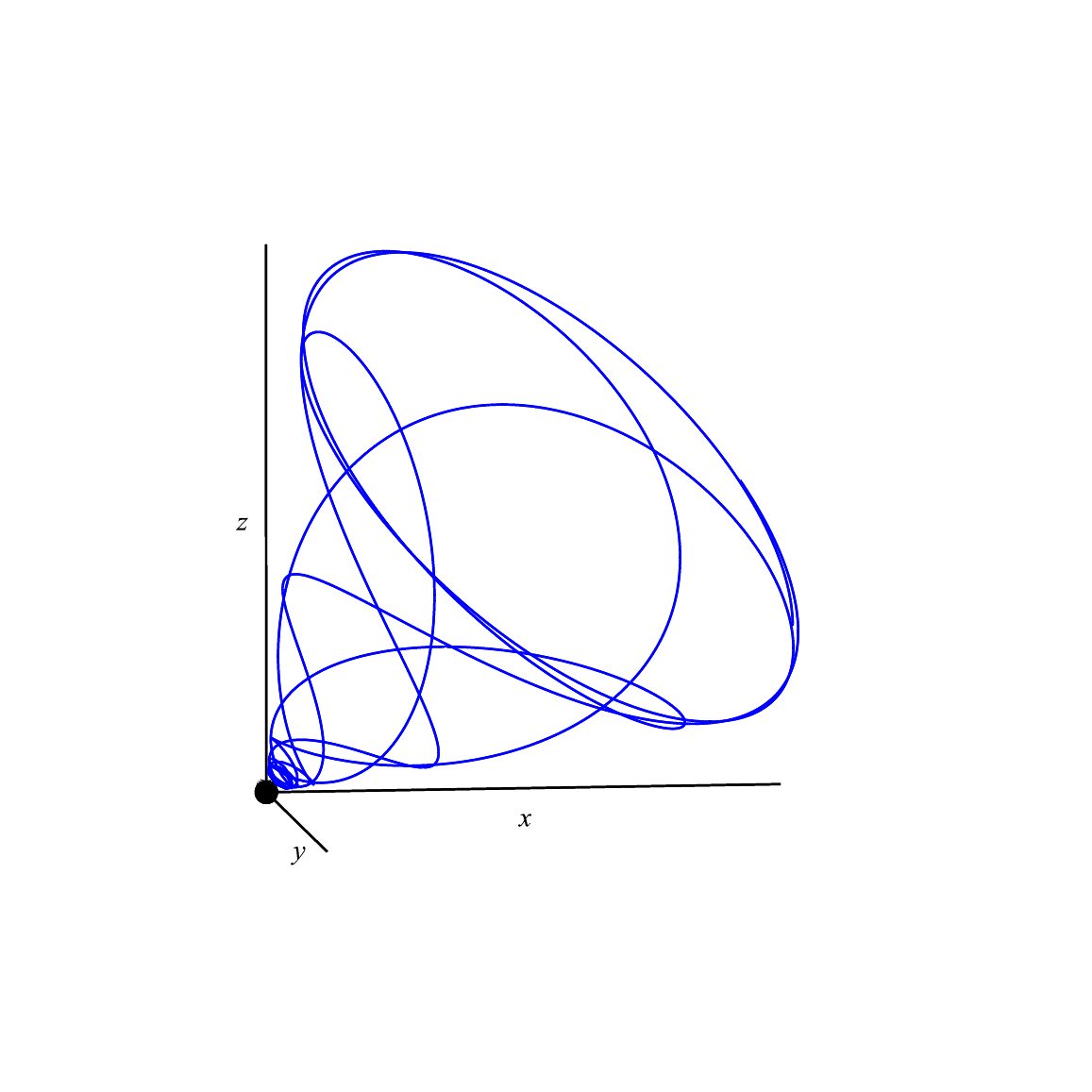}
\end{center}
\caption{
 {\it Left panel}:
The first unstable circular orbit for a charged particle with $\varepsilon Q_m=3$, $L=1.6$ and energy $E_1^2=0.9988$. The particle escapes towards the cosmological horizon.
 {\it Right panel}:
The second unstable circular orbit for a particle with $\varepsilon Q_m=3$, $L=0.5$ and energy $E_3^2=0.9666$. The particle moves into a bound orbit around the black hole 
}
\label{fig5}
\end{figure}

The first unstable circular orbit corresponds to the energy value $E_1^2=0.9988$, while the second unstable circular orbit has $E_2^2=0.9666$. Finally, the stable circular orbit has the energy $E_3^2=0.9239$. As such, if the charged particle has energy greater than $E_1$ then it either falls into the black hole or escapes towards the cosmological horizon, depending on the chosen initial conditions. In Figure \ref{fig5}, in the left panel we plot the first unstable circular orbit, with the charged particle escaping towards the cosmological horizon. In the right panel we show the trajectory of a charged particle orbiting the stable circular orbit.

\begin{figure}[ht!]
\begin{center} 
\includegraphics[width=0.43\textwidth, angle =0 ]{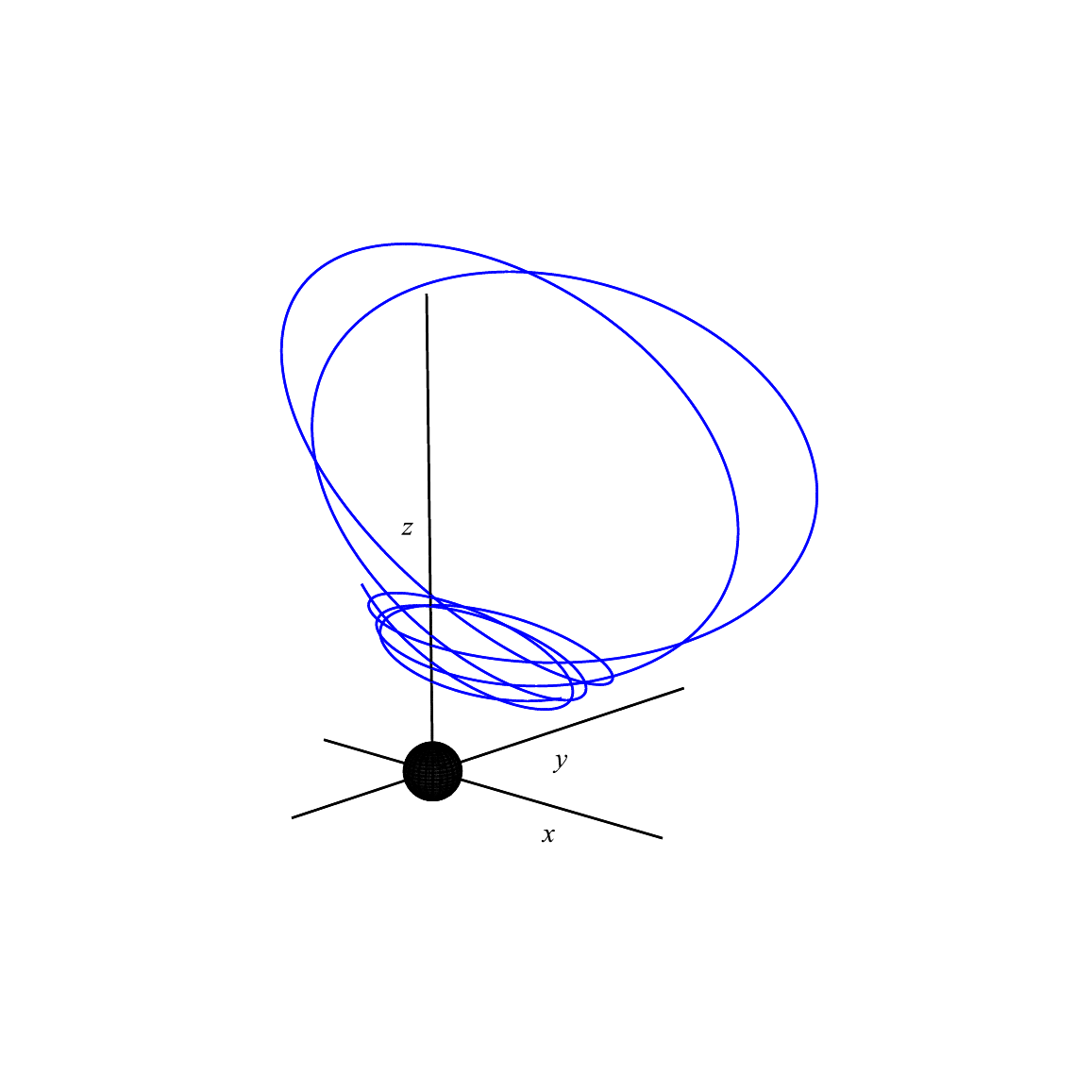}
\includegraphics[width=0.43\textwidth, angle =0 ]{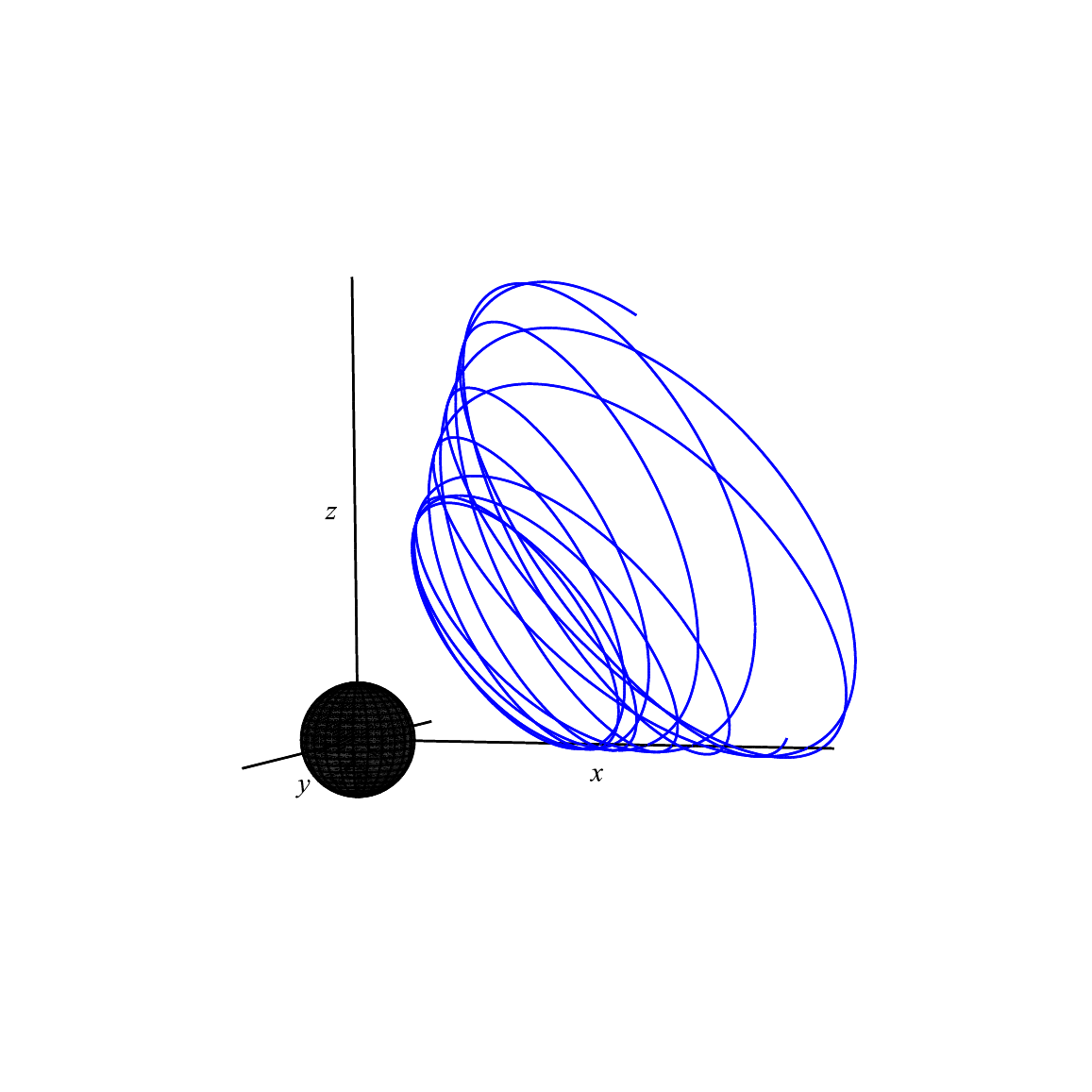}
\end{center}
\caption{
 {\it Left panel}:
The bounded trajectory of a charged particle with $\varepsilon Q_m=3$, $L=-0.5$ and energy $E^2=0.95$.
 {\it Right panel}:
The bounded trajectory for a particle with $\varepsilon Q_m=3$, $L=1$ and energy $E^2=0.93$.
}
\label{fig6}
\end{figure}

In Figure \ref{fig6} we show two bounded trajectories for a particle with energies between $E_3<E<E_2$.

Note that the motion is confined to Poincar\'e cones of various angles and inclinations.

\section{Conclusions}

In recent years, Kiselev's solution has received increased interest in connection to the properties of the anisotropic fluid sourcing this geometry, properties that mimic a dark energy source. In the present work we use a reinterpretation of the Kiselev geometry in the context of nonlinear electrodynamics theories. More specifically, we focused on the solution presented in \cite{Dariescu:2022kof}. This solution can be sourced either by an electric charge or by a magnetic charge, in the power-Maxwell theory. The power-Maxwell theory was usually studied in the asymptotically flat regime, when the power $p$ is negative. In our work we pushed this regime to extreme, by realizing that a geometry like that of Kiselev (or like the de Sitter geometry) can also be sourced by nonlinear electromagnetic fields in the power-Maxwell theory. However, when a nonlinear Maxwell field sources these geometries there are some issues with the photon propagation in these backgrounds, since in this case the photons will not move on null geodesics of the original geometry. Instead, they will move on an effective geometry as discussed in \cite{Dariescu:2022kof} and the problem arises here since that effective geometry has the wrong signature. Therefore, an easy point to confront the theory with experimental data is not available: there is no easy way to construct black hole images in this theory, which might signal a pathology of some sort.

The purpose of this paper was to present a detailed characterization of the possible motions of electrically charged test particles in these backgrounds. In Section $3$ we investigated the influence of the nonlinear electric field on the motion of a electrically charged particle. As it turns out, the value of the charge $Q_e$ has a deep influence on the shape of the effective potential and also on the location of the horizons, which tend to approach each other as $Q_e$ is increasing. As it can be noticed in Figure \ref{VFig}, the motion of particles with $V_{min} < E < V_{max}$ can be bounded outside the black hole horizon.

In Section $4$ we investigated the motion of electrically charged particles around a magnetically charged Kiselev black hole in the power-Maxwell theory. In this case the motion is confined on Poincar\'e cones of various angles. 

As avenues for further work, it might be interesting to investigate the effects of nonlinear power-Maxwell fields in constructing compact objects in General Relativity. In the usual Maxwell theory such solutions were generated for instance in \cite{Stelea:2018cgm}, \cite{Stelea:2018elx} and it might be fruitful to further investigate this matter in the more general context of nonlinear electrodynamics. 

Another interesting issue is the study of the behavior of charged scalar and spinorial fields in the background of the reinterpreted Kiselev geometry in the nonlinear electrodynamics. Following similar analysis performed in \cite{Dariescu:2019psb} - \cite{Dariescu:2018dyy}, it is quite possible that for particular values of the parameter $p$ the solutions can be expressed analytically by means of the Heun functions \cite{Birkandan:2017rdp}. It might also prove fruitful to rewrite the Kiselev geometry using inflationary coordinates as in \cite{Brihaye:2006kn} and look for multi-black holes in this context or in the Kaluza-Klein geometries in $5$ dimensions \cite{Stelea:2009ur}. Another interesting work might be related to treating the power-Maxwell electromagntic field as a dark matter/energy field, besides the usual Maxwell field, along the work done in \cite{Morris:2023vhv}.
 Work on these issues is in progress and it will be reported elsewhere.

\section*{Acknowledgements}

The authors would like to thank the anonymous referees whose remarks and suggestions helped improve this manuscript.


\begin{thebibliography}{99}

\bibitem{Riess}
A.~G.~Riess {\it et al.},
Astron.\ J.\  {\bf 116}, 1009 (1998).

\bibitem{Perlmutter}
S.~Perlmutter {\it et al.},
Astrophys.\ J.\  {\bf 517}, 565 (1999).

\bibitem{WMAP1} 
D.~N.~Spergel {\it et al.}  [WMAP Collaboration],
Astrophys.\ J.\ Suppl.\  {\bf 148}, 175 (2003).
 
\bibitem{Planck} 
P.~A.~R.~Ade {\it et al.}  [Planck Collaboration],
arXiv:1303.5076 [astro-ph.CO].
 
\bibitem{BAO1}
D.~J.~Eisenstein {\it et al.}  [SDSS Collaboration],
Astrophys.\ J.\  {\bf 633}, 560 (2005).

\bibitem{ORaifeartaigh:2017uct}
C.~O'Raifeartaigh, M.~O'Keeffe, W.~Nahm and S.~Mitton,
Eur. Phys. J. H \textbf{42}, no.3, 431-474 (2017)
doi:10.1140/epjh/e2017-80002-5
[arXiv:1701.07261 [physics.hist-ph]].

\bibitem{Padmanabhan:2002ji}
T.~Padmanabhan,
Phys. Rept. \textbf{380}, 235-320 (2003)
doi:10.1016/S0370-1573(03)00120-0
[arXiv:hep-th/0212290 [hep-th]].

\bibitem{Capozziello:2005ra}
S.~Capozziello, V.~F.~Cardone, E.~Piedipalumbo and C.~Rubano,
Class. Quant. Grav. \textbf{23}, 1205-1216 (2006)
doi:10.1088/0264-9381/23/4/009
[arXiv:astro-ph/0507438 [astro-ph]].

\bibitem{Vikman:2004dc}
A.~Vikman,
Phys. Rev. D \textbf{71}, 023515 (2005)
doi:10.1103/PhysRevD.71.023515
[arXiv:astro-ph/0407107 [astro-ph]].

\bibitem{Tsujikawa:2013fta}
S.~Tsujikawa,
Class. Quant. Grav. \textbf{30}, 214003 (2013)
doi:10.1088/0264-9381/30/21/214003
[arXiv:1304.1961 [gr-qc]].

\bibitem{Wolf:2023uno}
W.~J.~Wolf and P.~G.~Ferreira,
Phys. Rev. D \textbf{108}, no.10, 103519 (2023)
doi:10.1103/PhysRevD.108.103519
[arXiv:2310.07482 [astro-ph.CO]].

\bibitem{Capozziello:2005mj}
S.~Capozziello, S.~Nojiri and S.~D.~Odintsov,
Phys. Lett. B \textbf{634}, 93-100 (2006)
doi:10.1016/j.physletb.2006.01.065
[arXiv:hep-th/0512118 [hep-th]].


\bibitem{Kiselev:2002dx}
V.~V.~Kiselev,
Class. Quant. Grav. \textbf{20}, 1187-1198 (2003)
doi:10.1088/0264-9381/20/6/310
[arXiv:gr-qc/0210040 [gr-qc]].

\bibitem{Visser:2019brz}
M.~Visser,
Class. Quant. Grav. \textbf{37}, no.4, 045001 (2020)
doi:10.1088/1361-6382/ab60b8
[arXiv:1908.11058 [gr-qc]].

\bibitem{Dariescu:2022kof}
M.~A.~Dariescu, C.~Dariescu, V.~Lungu and C.~Stelea,
Phys. Rev. D \textbf{106}, no.6, 064017 (2022)
doi:10.1103/PhysRevD.106.064017
[arXiv:2206.12876 [gr-qc]].

\bibitem{Born}
Born, M. Nature 132, 282 (1933). https://doi.org/10.1038/132282a0,
Born, M., Infeld, L. Nature 132, 1004 (1933). https://doi.org/10.1038/1321004b0

\bibitem{Euler} 
Heisenberg, W. \& Euler, H.\ 2006, physics/0605038


\bibitem{Breton:2007bza}
N.~Breton and R.~Garcia-Salcedo,
[arXiv:hep-th/0702008 [hep-th]].

\bibitem{Bokulic:2021dtz}
A.~Bokuli\'c, T.~Juri\'c and I.~Smoli\'c,
Phys. Rev. D \textbf{103}, no.12, 124059 (2021)
doi:10.1103/PhysRevD.103.124059
[arXiv:2102.06213 [gr-qc]].

\bibitem{Hassaine:2008pw}
M.~Hassaine and C.~Martinez,
Class. Quant. Grav. \textbf{25}, 195023 (2008)
doi:10.1088/0264-9381/25/19/195023
[arXiv:0803.2946 [hep-th]].

\bibitem{Gonzalez:2009nn}
H.~A.~Gonzalez, M.~Hassaine and C.~Martinez,
Phys. Rev. D \textbf{80}, 104008 (2009)
doi:10.1103/PhysRevD.80.104008
[arXiv:0909.1365 [hep-th]].

\bibitem{Hassaine:2007py}
M.~Hassaine and C.~Martinez,
Phys. Rev. D \textbf{75}, 027502 (2007)
doi:10.1103/PhysRevD.75.027502
[arXiv:hep-th/0701058 [hep-th]].

\bibitem{EslamPanah:2021xaf}
B.~Eslam Panah,
EPL \textbf{134}, 20005 (2021)
doi:10.1209/0295-5075/134/20005
[arXiv:2103.08343 [physics.class-ph]].

\bibitem{Panah:2022cay}
B.~E.~Panah, K.~Jafarzade and A.~Rincon,
[arXiv:2201.13211 [physics.gen-ph]].

\bibitem{Hendi:2017lgb}
S.~H.~Hendi, B.~E.~Panah and S.~Panahiyan,
Fortsch. Phys. \textbf{66}, no.3, 1800005 (2018)
doi:10.1002/prop.201800005
[arXiv:1708.02239 [hep-th]].

\bibitem{Hendi:2017mgb}
S.~H.~Hendi, B.~Eslam Panah, S.~Panahiyan and A.~Sheykhi,
Phys. Lett. B \textbf{767}, 214-225 (2017)
doi:10.1016/j.physletb.2017.01.066
[arXiv:1703.03403 [gr-qc]].

\bibitem{Hendi:2016usw}
S.~H.~Hendi, B.~Eslam Panah, S.~Panahiyan and M.~S.~Talezadeh,
Eur. Phys. J. C \textbf{77}, no.2, 133 (2017)
doi:10.1140/epjc/s10052-017-4693-0
[arXiv:1612.00721 [hep-th]].

\bibitem{Hendi:2010zza}
S.~H.~Hendi and B.~E.~Panah,
Phys. Lett. B \textbf{684}, 77-84 (2010)
doi:10.1016/j.physletb.2010.01.026
[arXiv:1008.0102 [hep-th]].

\bibitem{Jeong:2023hom}
S.~Jeong, B.~H.~Lee, H.~Lee and W.~Lee,
Phys. Rev. D \textbf{107}, no.10, 104037 (2023)
doi:10.1103/PhysRevD.107.104037
[arXiv:2301.12198 [gr-qc]].

\bibitem{Poincare}
H.~Poincar\'e,
Compt. Rendus 123 (1896) 530–533.

\bibitem{Lim:2022qrt}
Y.~K.~Lim,
Phys. Rev. D \textbf{106}, no.6, 064023 (2022)
doi:10.1103/PhysRevD.106.064023
[arXiv:2206.00170 [gr-qc]].

\bibitem{Lim:2021ejg}
Y.~K.~Lim,
Phys. Rev. D \textbf{103}, no.8, 084044 (2021)
doi:10.1103/PhysRevD.103.084044
[arXiv:2102.08531 [gr-qc]].

\bibitem{Pugliese:2011py}
D.~Pugliese, H.~Quevedo and R.~Ruffini,
Phys. Rev. D \textbf{83}, 104052 (2011)
doi:10.1103/PhysRevD.83.104052
[arXiv:1103.1807 [gr-qc]].

\bibitem{Grunau:2010gd}
S.~Grunau and V.~Kagramanova,
Phys. Rev. D \textbf{83}, 044009 (2011)
doi:10.1103/PhysRevD.83.044009
[arXiv:1011.5399 [gr-qc]].

\bibitem{Pugliese:2011py}
D.~Pugliese, H.~Quevedo and R.~Ruffini,
Phys. Rev. D \textbf{83}, 104052 (2011)
doi:10.1103/PhysRevD.83.104052
[arXiv:1103.1807 [gr-qc]].

\bibitem{Mino:2003yg}
Y.~Mino,
Phys. Rev. D \textbf{67}, 084027 (2003)
doi:10.1103/PhysRevD.67.084027
[arXiv:gr-qc/0302075 [gr-qc]].

\bibitem{Baines:2021qaw}
J.~Baines, T.~Berry, A.~Simpson and M.~Visser,
Universe \textbf{7}, no.12, 473 (2021)
doi:10.3390/universe7120473
[arXiv:2110.01814 [gr-qc]].

\bibitem{Fathi:2022pqv}
M.~Fathi, M.~Olivares and J.~R.~Villanueva,
Eur. Phys. J. C \textbf{82}, no.7, 629 (2022)
doi:10.1140/epjc/s10052-022-10600-w
[arXiv:2205.13261 [gr-qc]].

\bibitem{Wang:2023otn}
R.~Wang, F.~Gao and H.~Chen,
Phys. Dark Univ. \textbf{40}, 101189 (2023)
doi:10.1016/j.dark.2023.101189

\bibitem{Stelea:2018cgm}
C.~Stelea, M.~A.~Dariescu and C.~Dariescu,
Phys. Rev. D \textbf{97} (2018) no.10, 104059
doi:10.1103/PhysRevD.97.104059
[arXiv:1804.08075 [gr-qc]].

\bibitem{Stelea:2018elx}
C.~Stelea, M.~A.~Dariescu and C.~Dariescu,
Phys. Rev. D \textbf{108}, no.8, 084034 (2023)
doi:10.1103/PhysRevD.108.084034
[arXiv:1810.02235 [gr-qc]].

\bibitem{Dariescu:2019psb}
M.~A.~Dariescu, C.~Dariescu and C.~Stelea,
Mod. Phys. Lett. A \textbf{35}, no.07, 2050036 (2019)
doi:10.1142/S0217732320500364
[arXiv:1903.03552 [hep-th]].

\bibitem{Dariescu:2017ima}
C.~Dariescu, M.~A.~Dariescu and C.~Stelea,
Gen. Rel. Grav. \textbf{49}, no.12, 153 (2017)
doi:10.1007/s10714-017-2314-8

\bibitem{Dariescu:2018dyy}
M.~A.~Dariescu, C.~Dariescu and C.~Stelea,
Gen. Rel. Grav. \textbf{50}, no.10, 126 (2018)
doi:10.1007/s10714-018-2449-2

\bibitem{Birkandan:2017rdp}
T.~Birkandan and M.~Horta\c{c}su,
EPL \textbf{119}, no.2, 20002 (2017)
doi:10.1209/0295-5075/119/20002
[arXiv:1704.00294 [math-ph]].

\bibitem{Brihaye:2006kn}
Y.~Brihaye, B.~Hartmann, E.~Radu and C.~Stelea,
Nucl. Phys. B \textbf{763}, 115-146 (2007)
doi:10.1016/j.nuclphysb.2006.11.011
[arXiv:gr-qc/0607078 [gr-qc]].

\bibitem{Stelea:2009ur}
C.~Stelea, K.~Schleich and D.~Witt,
Phys. Rev. D \textbf{83}, 084037 (2011)
doi:10.1103/PhysRevD.83.084037
[arXiv:0909.3835 [hep-th]].

\bibitem{Morris:2023vhv}
J.~R.~Morris,
Phys. Lett. B \textbf{847}, 138325 (2023)
doi:10.1016/j.physletb.2023.138325
[arXiv:2311.10890 [gr-qc]].

\end{thebibliography}
\end{document}